\documentclass{iopart}
\usepackage{iopams}
\usepackage{graphicx}
\usepackage[colorlinks=true,citecolor=blue,linkcolor=blue,pdfstartview=FitH,pdfauthor=Niemczyk]{hyperref}

\begin{document}
\article[Superconducting Quantum Circuits]{}
      {Fabrication Technology of and Symmetry Breaking in Superconducting Quantum Circuits}

\author{T. Niemczyk$^{1}$, F. Deppe$^{1,2}$, M. Mariantoni$^{1,2}$, E.P. Menzel$^{1}$, E. Hoffmann$^{1,2}$,
G. Wild$^{1,2}$, L. Eggenstein$^{1,2}$, A. Marx$^{1}$, R. Gross$^{1,2}$}

\address{$^{1}$Walther-Mei{\ss}ner-Institut, Bayerische Akademie der Wissenschaften,
        D-85748 Garching, Germany}
\address{$^2$Physik-Department, Technische Universit\"{a}t M\"{u}nchen, D-85748 Garching,
        Germany }
\ead{Rudolf.Gross@wmi.badw.de}

\begin{abstract}
Superconducting quantum circuits are promising systems for experiments testing
fundamental quantum mechanics on a macroscopic scale and for applications in
quantum information processing. We report on the fabrication and
characterization of superconducting flux qubits, readout dc SQUIDs, on-chip
shunting capacitors, and high-quality coplanar waveguide resonators.
Furthermore, we discuss the tunability and fundamental symmetry aspects
inherent to all superconducting qubits, which can be regarded as artificial
solid-state atoms. Comparing them to their natural counterparts, we discuss
first and second-order energy shifts due to static control fields.
Additionally, we present an intuitive derivation of the first- and second-order
matrix elements for level transitions in the presence of a coherent microwave
driving.
\end{abstract}

\pacs{85.25.-j, 
      85.25.Cp  
      03.67.–a  
     }

\submitto{\SUST}

\section{Introduction}

Over the last decade the idea of solid-state based quantum information
processing has excited scientists and engineers in many disciplines, seeking to
combine two major developments of the last century, namely quantum mechanics
and integrated circuit technology~\cite{Nielsen:2000a}. Although the practical
challenges to realize solid-state based quantum information circuits are huge,
the basic building blocks have been successfully implemented. Nevertheless,
their fabrication and measurement remain demanding tasks. At present, most
experiments are addressing two key questions.  First, how can we fabricate,
control, and read-out solid-state quantum circuits with sufficiently long
coherence times, and second, how can we establish a controlled coupling between
distant circuit parts and transfer the quantum information between them.
Regarding the first problem, we focus on the fabrication of several key
elements of superconducting quantum circuits. Their quantum coherent dynamics
was discussed in our previous
work~\cite{Deppe:2008a,Kakuyanagi:2007a,Deppe:2007a}. With respect to the
second one, we address symmetry aspects of these circuits. In particular, we
discuss how symmetry can be broken in a controlled way using external control
parameters~\cite{Deppe:2008a}.

In this context, superconducting circuits containing nonlinear elements such as
Josephson junctions or phase-slip centers play an important role. The lateral
dimensions of these circuits and their constituents can range from 100\,nm to a
few hundreds of $\mu$m. Such circuits constitute artificial atoms in the sense
that they behave similarly to natural atoms in many
aspects~\cite{Clarke:2008a}.  Despite the fact that these artificial atoms are
huge compared to their natural counterparts, they have a discrete level
structure and exhibit properties unique to the world of quantum mechanics. In
the simplest case, they form quantum two-level systems, also referred to as
quantum bits or qubits. Such qubits are particularly attractive for the
investigation of fundamental quantum phenomena on a macroscopic scale as well
as for the use in quantum information systems. A major advantage of artificial
solid-state atoms over natural atoms is their design flexibility and wide
tunability by means of external parameters such as electric or magnetic fields. 	
Coupling superconducting qubits to on-chip microwave resonators has given rise
to circuit quantum electrodynamics
(QED)~\cite{Schoelkopf:2008a,Wallraff:2004a,Chiorescu:2004a,Blais:2004a,Mariantoni:2005a,Johansson:2006a,Houck:2007a,Sillanpaa:2007a,Majer:2007a,Astafiev:2007a,Hofheinz:2008a,Fink:2008a,Mariantoni:2008a}.
In this prospering field, the fundamental interaction between artificial
solid-state atoms (qubits) and (single) microwave photons is studied. We note
that circuit QED is the solid-state analog of cavity
QED~\cite{Mabuchi:2002,Haroche:2006a,Walther:2006a} in quantum optics, where
natural atoms are coupled to suitable three-dimensional cavities. However, in
contrast to quantum-optical cavity QED, the strong coupling regime, where the
qubit-photon interaction coefficient is much larger than all relevant decay
rates of qubit and resonator, can be achieved easily in circuit
QED~\cite{Wallraff:2004a,Chiorescu:2004a}. Moreover, solid-state circuits
exhibit an inherent tunability. We have recently shown that this can be
exploited to break the symmetry of a superconducting flux qubit coupled to a
lumped-element microwave resonator in a controlled way. This allowed us to
study multi-photon excitations and to get fundamental insight into the
interplay of multi-photon processes and symmetries in a qubit-resonator
system~\cite{Deppe:2008a}. In another work, we introduced a systematic
formalism for two-resonator circuit~QED, where two on-chip microwave resonators
are simultaneously coupled to one superconducting
qubit~\cite{Mariantoni:2008a}. Within this framework, we demonstrated that the
qubit can function as a quantum switch mediating a tunable interaction between
the two resonators, which are assumed to be originally independent
\cite{Mariantoni:2008a}.

This article is composed as follows. In section~\ref{Fabrication of
Superconducting Quantum Circuits}, we focus on the fabrication of
superconducting flux qubits, their readout circuitry and high-quality
superconducting on-chip resonators. Then, in section~\ref{Tunability Coupling
Strength Symmetry Breaking}, we address fundamental symmetry aspects of
artificial solid-state atoms and compare them to their natural counterparts. As
it turns out, these results hold for a general quantum two-level
system~\cite{Cohen-Tannoudji:1977}. In particular, we spell out the expressions
for both flux and charge qubits.

\section{Fabrication of Superconducting Quantum Circuits}
\label{Fabrication of Superconducting Quantum Circuits}

Superconducting quantum circuits are based on flux quantization and Josephson
tunneling. They have been studied intensively because they have opened a new
area in fundamental science and because of their potential for the realization
of quantum information processing systems \cite{Clarke:2008a}. Here, we discuss
the fabrication process of superconducting three-junction flux qubits and the
additional circuit elements required for readout, coupling and manipulation.

\subsection{Flux Qubits}
\label{Flux Qubits}

Today, there are three fundamental types of superconducting qubits, namely
flux, charge and phase qubits \cite{Clarke:2008a}, depending on the nature of
the relevant quantum variable. The three-junction flux qubit
\cite{Mooij:1999a,Orlando:1999a,van der Wal:2000a}, we are focussing on here,
consists of a superconducting loop interrupted by three nm-sized Josephson
junctions. Whereas two of these junctions with area $A$ and critical current
$I_{\rm c}$ are equal, one junction is designed to have smaller area $\alpha A$
and critical current $\alpha I_{\rm c}$ with $\alpha \sim 0.6-0.8$. In an
equivalent circuit the Josephson junctions are represented by a normal
resistance $R_{\rm n}$, a junction capacitance $C_{\rm J}$ and a Josephson
inductance $L_{\rm J} = L_{\rm c}/\cos\varphi$ with $L_{\rm c} = \hbar/2eI_{\rm
c}$. Here, $\varphi$ is the phase difference across the junction, $\hbar$ the
reduced Planck's constant, and $e$ the elementary charge. The characteristic
energies are the Josephson coupling energy of the junctions, $E_{\rm J} = \hbar
I_{\rm c}/2e$, associated with the storage of a single flux quantum $\Phi_0
=h/2e$ in the Josephson inductor, and the charging energy, $E_{\rm c} =
e^2/2C_{\rm J}$, associated with the storage of a single electron charge $e$ on
the junction capacitance $C_{\rm J}$. For flux qubits, $E_{\rm J} \gg E_{\rm
c}$ (typically $10 \le E_{\rm J}/E_{\rm c} \le 100$) to ensure that the
magnetic flux $\Phi$ in the loop is the relevant quantum variable and the
ground and the first excited state of the qubit have opposite circulating
currents which produce measurable flux. The minimum temperature for experiments
on superconducting qubits usually is limited to the base temperature $T_{\rm b}
\gtrsim 10$\,mK of dilution refrigerators. To avoid thermal population of the
upper qubit state, the level splitting $\Delta_{\rm ge}$ between the ground
state $|\textrm{g} \rangle$ and the first excited state $|\textrm{e} \rangle$
should be much larger than $k_{\rm B}T_{\rm b}$, corresponding to an energy and
frequency of about $1\,\mu$eV and 200\,MHz, respectively, at $T_{\rm b}
=10$\,mK. Hence, $\Delta_{\rm ge}/h \gtrsim 2$\,GHz is desired. Since
$\Delta_{\rm ge} \propto \hbar\omega_{\rm p} \exp(-a\sqrt{E_{\rm J}/E_{\rm
c}})$, where $\omega_{\rm p} \propto E_{\rm c}\sqrt{E_{\rm J}/E_{\rm c}}$ is
the plasma frequency and $a$ a constant of the order of unity, large $E_{\rm
c}$ is required \cite{Mooij:1999a,Orlando:1999a}. At the same time one has to
keep $E_{\rm J}/E_{\rm c}$ large enough to have well-defined flux states with a
measurable circulating current. These requirements are demanding regarding
junction fabrication technology, asking for junctions with high critical current densities $J_{\rm c}
\simeq 10^3$A/cm$^2$ and areas down to a few $0.01\,\mu\textrm{m}^2$. Since
such small junctions with small parameter spread are difficult to fabricate by
the well established Nb/AlO$_x$/Nb technology, most flux qubits so far are
based on Al/AlO$_x$/Al junctions fabricated by electron beam lithography and
two-angle shadow evaporation \cite{Dolan:1977}.

\begin{figure}[tb]
 \centering{
    \includegraphics[width=0.85\columnwidth]{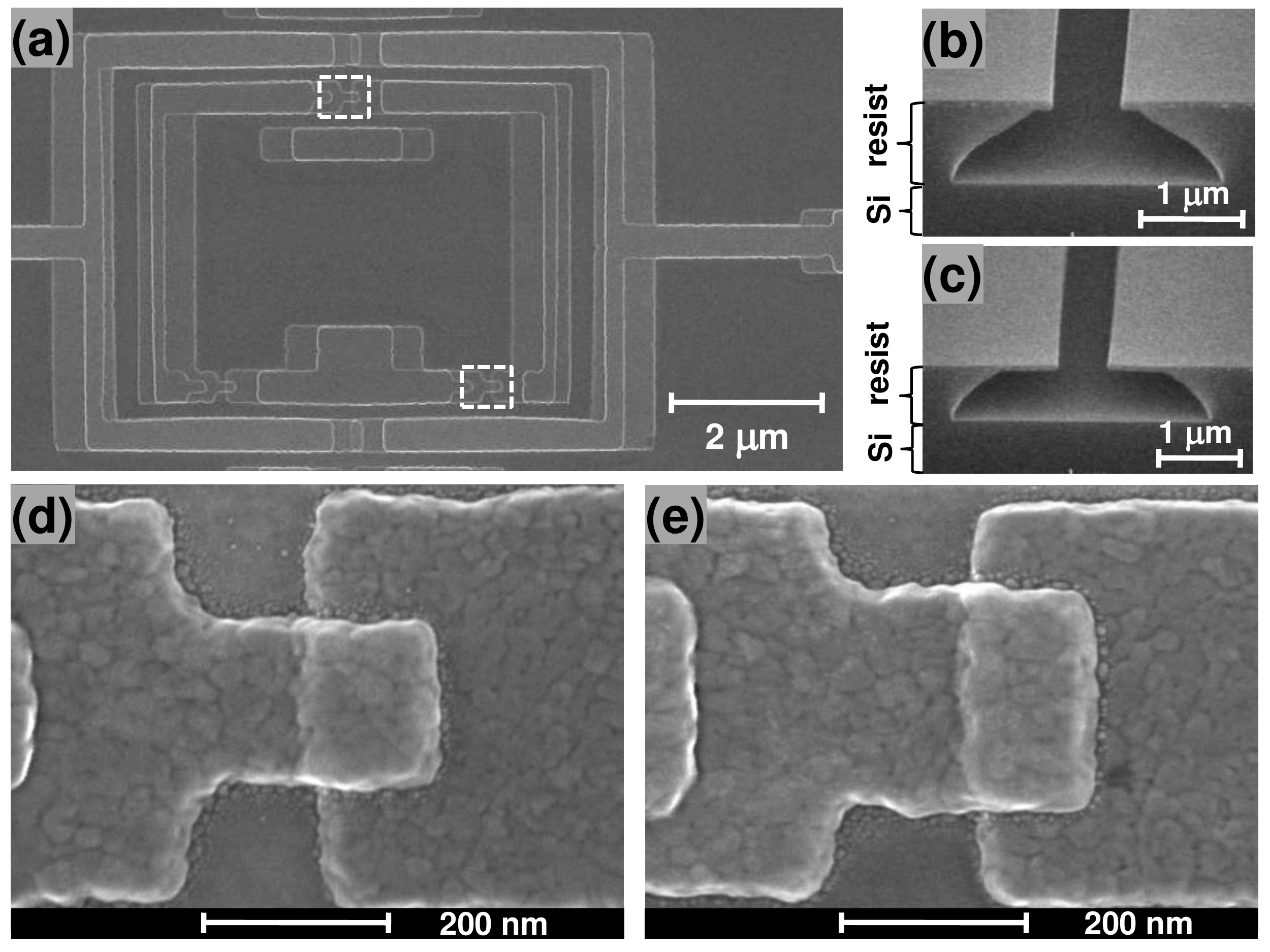}}
    \caption{(a) SEM micrograph of a three-junction flux qubit (inner loop)
             surrounded by the readout dc SQUID (outer loop with two junctions). The white rectangles mark the
             regions of the $\alpha$-junction (upper) with a reduced area ($\alpha = 0.7$) and a regular junction (lower)
             shown on an enlarged scale in (d) and (e), respectively.
             (b) and (c) show cross-sectional views of double-layer resist structures obtained with different
             exposure times. Large undercuts can be obtained, resulting in free-standing resist parts required
             for the shadow evaporation technique.
             }
    \label{fig:fluxqubit}
\end{figure}

Figure~\ref{fig:fluxqubit}(a) shows a scanning electron microscopy (SEM)
micrograph of a three-junction flux qubit based on Al/AlO$_x$/Al junctions. The
qubit is surrounded by the readout dc SQUID (Superconducting Quantum
Interference Device). Figures~\ref{fig:fluxqubit}(d) and (e) show an enlarged
view of the so-called $\alpha$-junction, which has a reduced area $\alpha A$
with $\alpha = 0.7$, and one of the regular qubit junctions, respectively. The
area of the $\alpha$-junction is only about $0.02\,\mu\textrm{m}^2$. The qubits
are fabricated on thermally oxidized (50\,nm SiO$_2$) Si wafers by electron
beam lithography using a Philips XL30~SFEG field emission SEM and a Raith Elphy
Plus nanolithography system. For producing free-standing resist masks we used a
double-layer resist system. Typical cross-sectional views of the resist
stencils are shown in Figs.~\ref{fig:fluxqubit}(b) and (c). Due to the larger
sensitivity of the underlay resist, large undercuts can be generated in a
controlled way. The two-angle shadow evaporation of Al was done by electron
beam evaporation in a UHV system with a base pressure in the $10^{-9}$mbar
range. The bottom layer was thermally oxidized \textit{in situ} in pure oxygen
at $p=2\times 10^{-4}$ mbar. In order to achieve high current densities we used
small products $L$ of oxygen pressure and oxidation time ranging between 0.25
and $0.26\,\textrm{mbar}\cdot \textrm{s}$. With these values we achieved
$J_{\rm c}(50\,\textrm{mK})\simeq 1500\pm 500$\,A/cm$^2$ with a high yield of
85\%. By optimizing the electron beam lithography process we could reduce the
spread in the junction area down to 10\%. We note that we can increase or
decrease $J_{\rm c}$ by decreasing or increasing $L$, respectively. For
example, by increasing $L$ to above $10\,\textrm{mbar}\cdot \textrm{s}$ the
$J_{\rm c}$ values can be reduced to below 10\,A/cm$^2$.

\begin{figure}[tb]
 \centering{
    \includegraphics[width=0.6\columnwidth]{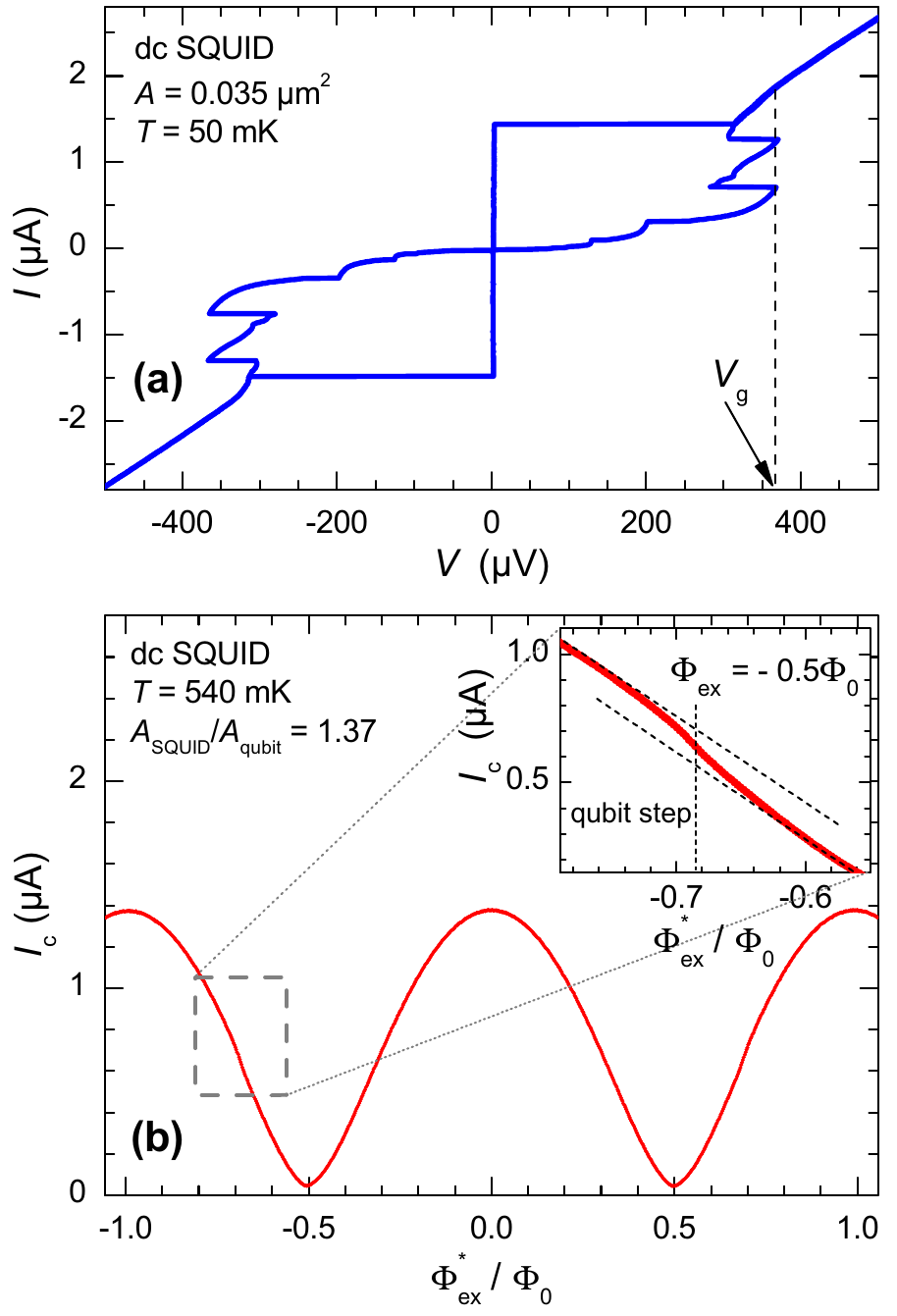}}
    \caption{(a) Current-voltage characteristics of a readout dc SQUID based on
             Al/AlO$_x$/Al tunnel junctions with $A=0.035\,\mu\textrm{m}^2$ at $T=50$\,mK.
             (b) Critical current $I_{\rm c}$ of a readout dc SQUID as a function of the applied
             magnetic flux $\Phi_{\rm ex}^\star$ threading the SQUID loop at $T=540$\,mK.
             The inset shows an enlarged view of the
             region around $\Phi_{\rm ex} = - \gamma \Phi_0/2$, where the qubit persistent current
             changes sign. Here, $\gamma = 1.37$ is the ratio of the loop areas of the SQUID and the qubit.
             }
    \label{fig:IVC}
\end{figure}

Figure~\ref{fig:IVC} shows the current-voltage characteristics (IVC) and the
flux dependence of the critical current of a readout dc SQUID fabricated with
the process described above. From $I_{\rm c}=1.44\,\mu\textrm{A}$ and the
measured junction area $A=0.035\,\mu\textrm{m}^2$ we obtain $J_{\rm c} =
2\,\textrm{kA/cm}^2$. Furthermore, from $R_{\rm n}=169\,\Omega$ the product
$I_{\rm c}R_{\rm n}= 243\,\mu\textrm{V}$ is obtained. The measured gap voltage
$V_{\rm g} \simeq 360\,\mu\textrm{V}$ is close to the BCS value $V_{\rm
g}=2\Delta/e=3.53 k_{\rm B}T_{\rm c}/e=365\,\mu\textrm{V}$ using $T_{\rm
c}=1.2\,\textrm{K}$ for the critical temperature of Al. Furthermore, the
measured $I_{\rm c}R_{\rm n}$ product agrees well with the Ambegaokar-Baratoff
value \cite{Ambegaokar:1963}  $\pi V_{\rm g}/4=287\,\mu\textrm{V}$. Using the
specific junction capacitance $C_{\rm s} = 100\pm
25\,\textrm{fF}/\mu\textrm{m}^2$ \cite{Deppe:2004}, the capacitance of the
qubit junctions with area $A=0.02\,\mu\textrm{m}^2$ is estimated to be $C_{\rm
J}\simeq 2\,\textrm{fF}$ corresponding to $E_{\rm c} = 40\,\mu\textrm{eV}$ or,
equivalently, $T=0.46\,$K. The subgap structures in the IVC in
Figure~\ref{fig:IVC}(a) most likely originate from $LC$ resonances as has been
discussed in more detail recently \cite{Deppe:2004}.

\begin{figure}[tb]
 \centering{
    \includegraphics[width=0.6\columnwidth]{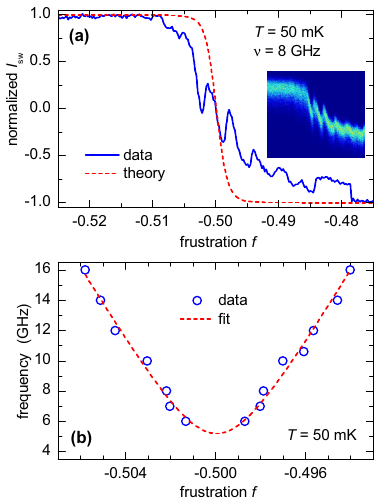}}
    \caption{(a) Normalized switching current $I_{\rm sw}$ of the readout dc SQUID plotted versus
             the frustration $f=\Phi_{\rm ex}/\Phi_0$ in the region around the qubit step at
             $T=50$\,mK. The qubit is irradiated by a microwave signal ($\nu = 8\,$GHz).
             The dashed curve shows the theoretically expected curve without microwave irradiation
             for $\Delta_{\rm ge}/h=5.2\,$GHz. The inset shows the switching current distribution in a
             color coded image.
             (b) Qubit transition frequency $E_{\rm ge}/h$ (symbols) plotted versus the frustration $f$.
             The line is a numerical fit to the data yielding $\Delta_{\rm ge}/h=5.2\,$GHz
             and $I_{\rm p} =450\,$nA.
             }
    \label{fig:spectroscopy}
\end{figure}

We note that the fabrication of sub-micron sized Josephson junctions with
two-angle shadow evaporation is very common since it is easy to use. However,
there are also drawbacks. In particular, the reproducibility and parameter
spread of the junctions will be finally limited by the non-planar junction
geometry, resulting in a not well defined junction area. The related
fluctuations in the critical current become an increasing problem when one is
going to more complex quantum circuits. However, more severe is critical
current noise due to impurities and trapping sites in the tunneling barrier
leading to decoherence. Therefore, in the long run a planar junction geometry
and the use of epitaxial electrode and barrier layers may be the optimum
choice.

The $I_{\rm c}(\Phi_{\rm ex}^\star)$ dependence of a readout dc SQUID is shown
in Fig.~\ref{fig:IVC}(b). The measured curve is close to the ideal $|
\cos(\pi\Phi_{\rm ex}^\star/\Phi_0)|$ dependence expected for a dc SQUID with
$\beta_{\rm L} =2\pi L_{\rm s}I_{\rm c}/\Phi_0=L_{\rm s}/L_{\rm c}\ll 1$, for
which the Josephson inductance $L_{\rm c}$ dominates the geometric inductance
$L_{\rm s}$. However, there are small deviations, originating from the
additional flux due to the persistent current $I_{\rm p}$ circulating clock- or
counterclockwise in the qubit loop. In the region close to the applied flux
$\Phi_{\rm ex}$ corresponding to $\Phi_0/2$ in the qubit loop, $I_{\rm p}$
changes sign. This results in a step-like feature superimposed on the regular
$I_{\rm c}(\Phi_{\rm ex}^\star)$ dependence. This so-called qubit step is shown
in the inset of Fig.~\ref{fig:IVC}(b). With the ratio of 1.37 between the SQUID
and the qubit loop we estimate $\Phi_{\rm ex} =-0.685\Phi_0$ in good agreement
with the experiment.

The qubit level structure has been investigated by microwave spectroscopy.
Figure~\ref{fig:spectroscopy}(a) shows the normalized switching current of the
readout SQUID around the qubit step plotted versus the frustration $f=\Phi_{\rm
ex}/\Phi_0$. Here, $\Phi_{\rm ex}$ is the applied flux threading the qubit
loop. Note that the switching of the SQUID from the zero-voltage into the
voltage state is by quantum tunneling and hence is a statistical process.
Therefore, the normalized switching current $I_{\rm sw}(f)$ is obtained from
the switching current distribution measured for every $f$. This distribution is
shown in a color-coded image in the inset of Fig.~\ref{fig:spectroscopy}(a).
The $I_{\rm sw}(f)$ curve shows a pronounced peak-dip structure due to resonant
transitions between the qubit levels induced by the applied microwave
irradiation with frequency $\nu = 8\,$GHz. Note that multi-photon transitions
are observed depending on the applied microwave power. The position of the
peaks/dips (one-photon transitions) as a function of frequency and frustration
is shown in Fig.~\ref{fig:spectroscopy}(b). The data follow the expected
$E_{\rm ge}=\sqrt{\epsilon^2+\Delta_{\rm ge}^2}$ dependence, where $\epsilon =
2I_{\rm p}\Phi_0(f+0.5)$ is the energy bias and $\Delta_{\rm ge}$ the level
splitting at the degeneracy point $f=-0.5$. Fitting the data yields $I_{\rm
p}=450\,$nA and a qubit gap of $\Delta_{\rm ge}/h = 5.2\,$GHz.

\subsection{Bias and Readout Circuit}

\begin{figure}[tb]
 \centering{
    \includegraphics[width=0.9\columnwidth]{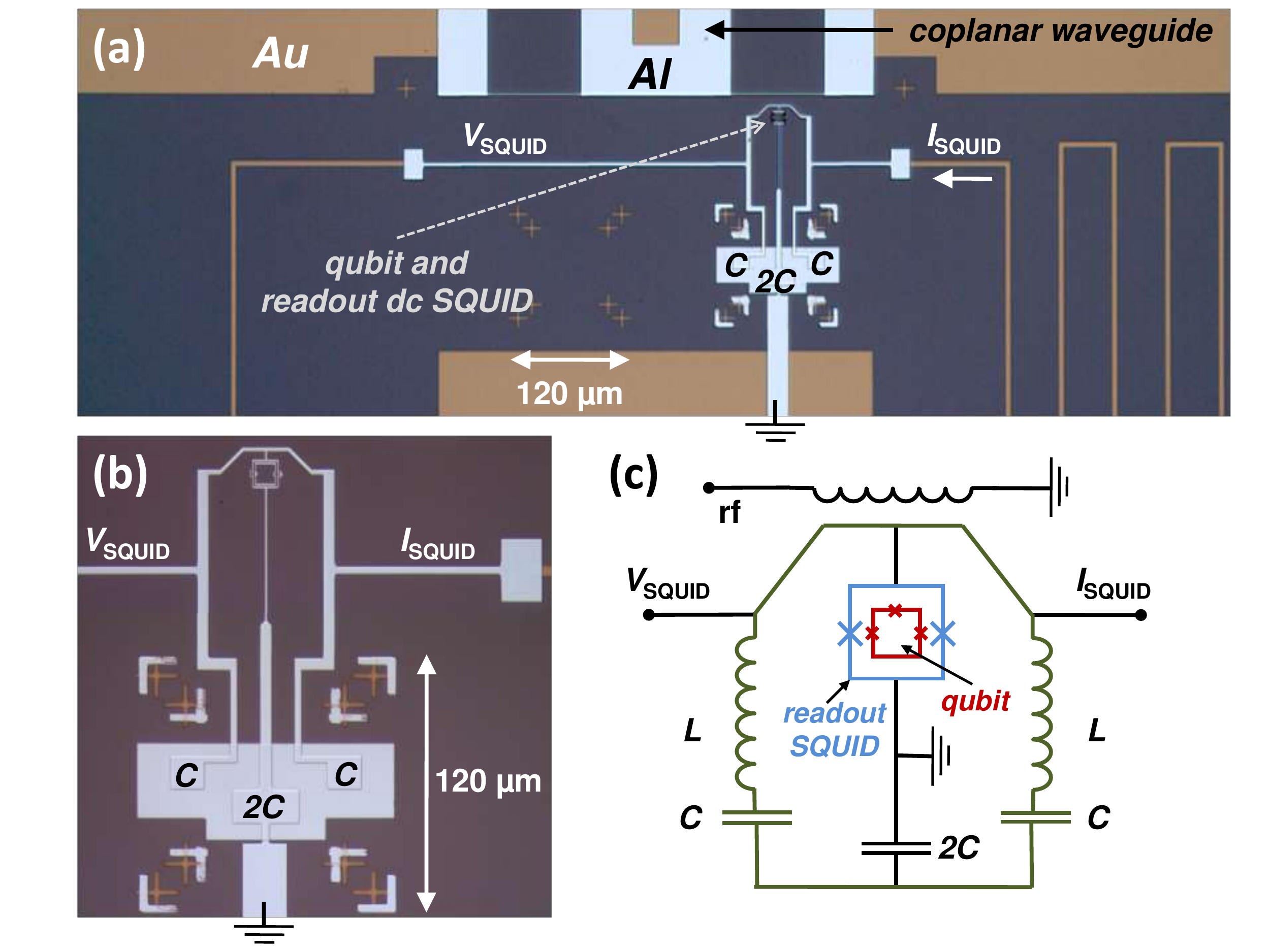}}
    \caption{(a) Optical micrograph of the readout and bias circuit of the dc SQUID and the
             end of the coplanar waveguide which is shorted to ground to form an antenna.
             In (b) an enlarged view of the region around the
             qubit and SQUID loop with the shunting capacitors is shown.
             The equivalent circuit is depicted in (c).
             }
    \label{fig:biascircuit}
\end{figure}

For the manipulation and readout of flux qubits as well as for the attenuation
of external electrical and magnetic noise additional circuit elements such as
resistors, capacitors, inductors, microwave lines, antennas etc.\ are required.
Some of them have to be fabricated on-chip. For the fabrication of these
usually larger circuit parts we used a mix\&match process, where high
resolution electron beam lithography patterns are precisely aligned into
existing patterns made by optical lithography. For the alignment markers, the
leads and the coplanar microwave lines we used 20\,nm thick Au layers. These
layers are fabricated by optical lithography, electron beam evaporation and a
final lift-off process prior to the e-beam process for the qubits described
above. During the subsequent Al evaporation process, parts of the Au pattern
are covered by Al. The small thickness of the Au layer allows to avoid the
interruption of the Al film at the edges of the Au layer.

Figures~\ref{fig:biascircuit}(a) and (b) show optical micrographs of the
circuit elements used for biasing the readout dc SQUID and measuring the SQUID
voltage. As shown by the equivalent circuit in Fig.~\ref{fig:biascircuit}(c),
the SQUID is shunted to ground by on-chip Al/AlO$_x$/Al capacitors with the
total capacitance $C$. In combination with the on-chip resistors in the dc
SQUID bias and voltage lines, which are realized by long Au lines, they form a
low-pass filter constituting the main component of the qubit electromagnetic
environment. Details on the shaping of the electromagnetic environment as well
as the effect of a resistive and capacitive coupling of the dc SQUID are given
elsewhere \cite{Deppe:2007a,Kakuyanagi:2007a}. The ground plane of the
Al/AlO$_x$/Al capacitors is deposited by electron beam evaporation prior to the
qubit fabrication process. It is thermally oxidized in ambient atmosphere for 3 hours at
$100^\circ$C to achieve a thick oxide layer with sufficiently small leakage
current. Specific capacitance values between 12 and
$15\,\textrm{fF}/\mu\textrm{m}^2$ and resistance times area products larger
than 100\,G$\Omega\mu$m$^2$ could be achieved.

It is evident from Fig.~\ref{fig:biascircuit}(c) that the two capacitors with
total capacitance $C/2$ together with the two inductors with total inductance
$2L$ form an $LC$ resonator coupled inductively to the qubit loop. The typical
resonance frequency $\nu_{\rm r} = \omega_{\rm r}/2\pi$ of the $LC$ resonator is in the GHz range
and is comparable to the qubit transition frequency. Furthermore, the coupling
strength $g/2\pi$ is of the order of 100\,MHz \cite{Deppe:2008a,Johansson:2006a}. Therefore,
the coupled qubit-resonator system allows to study fundamental questions of the
flourishing field of circuit QED
\cite{Wallraff:2004a,Chiorescu:2004a,Johansson:2006a,Houck:2007a,Sillanpaa:2007a,Majer:2007a,Astafiev:2007a,Wallraff:2007a}.
Here, the interaction of solid-state quantum systems with single microwave
photons is of particular interest. Recently, we have used a circuit very
similar to that shown in Fig.~\ref{fig:biascircuit} to observe one of the key
signatures of a two-photon driven Jaynes-Cummings model, which unveils the
upconversion dynamics of a superconducting flux qubit coupled to an on-chip
lumped-element microwave resonator \cite{Deppe:2008a}.

\subsection{Microwave Resonators}

The quality factor $Q_{\rm L}$ of the $LC$ resonators discussed above is only
of the order of 100. Therefore, the decay rate $\kappa = \nu_{\rm r}/Q_{\rm L}$
of the microwave photons from the resonator is large. In order to access the strong
coupling regime in circuit QED experiments, $g/2\pi \gg \kappa$ is required.
With $g/2\pi \sim 100\,$MHz and $\omega_{\rm r}/2\pi \sim 10\,$GHz, microwave
resonators with $Q_{\rm L} \gtrsim 10^4$ have to be designed. Such resonators
can be realized by simple transmission line geometries \cite{Frunzio:2005a}.

\begin{figure}[tb]
 \centering{
    \includegraphics[width=0.9\columnwidth]{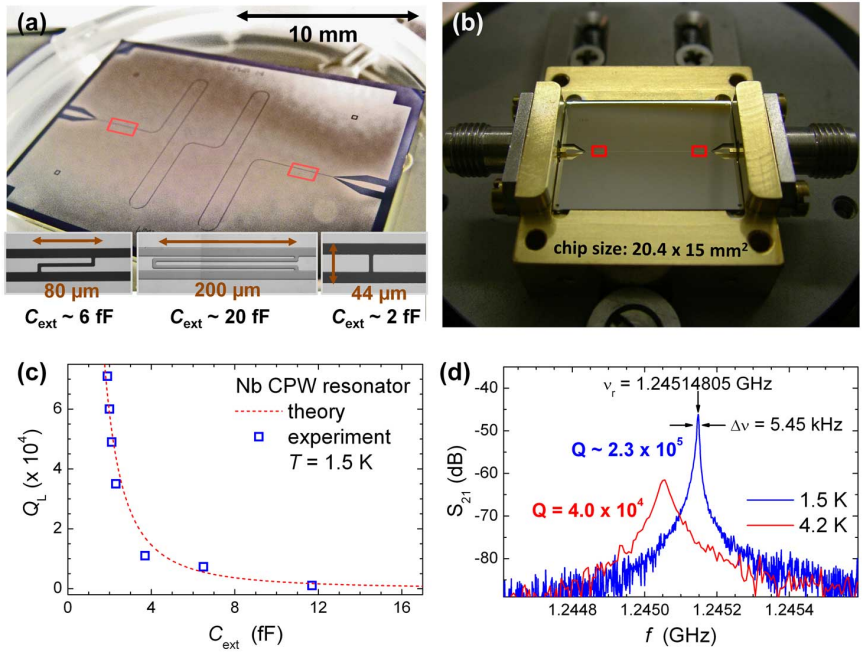}}
    \caption{(a) optical micrograph of a CPW microwave resonator. The insets show different coupling
             capacitors located at the positions marked by the rectangles.
             (b) Picture of the gold-plated copper sample box containing a resonator chip. The rectangles
             mark the positions of the coupling capacitors.
             (c) Measured $Q_{\rm L}$ (symbols) plotted versus the coupling capacitance for a 6\,GHz
             niobium CPW resonator
             at $T=1.5\,$K. The dashed line shows the theoretical prediction \cite{Pozar:2005a}.
             (d) Transmission spectrum of an undercoupled 1.25\,GHz niobium CPW resonator
             measured at 1.5 and 4.2\,K.
             }
    \label{fig:resonator}

\end{figure}

We have fabricated coplanar waveguide (CPW) resonators based on 200\,nm thick
Nb films on thermally oxidized Si and sapphire substrates. An optical
micrograph of such a resonator is shown in Fig.~\ref{fig:resonator}(a). The Nb
films have been deposited by dc magnetron sputtering in Ar. The patterning was
done by optical lithography and reactive ion etching using an Ar/SF$_6$
mixture. The center conductor of the CPW resonators is $20\,\mu\textrm{m}$
wide, separated from the lateral ground planes by a $12\,\mu\textrm{m}$ wide
gap to obtain a wave impedance of $50\,\Omega$. The resonator is coupled at
both ends to a coplanar transmission line by interdigital or gap capacitors
shown in the insets of Fig.~\ref{fig:resonator}(a). The strength of the
capacitive coupling determines the loaded quality factor $Q_{\rm L}^{-1} =
Q^{-1} + Q_{\rm ext}^{-1}$, where $Q$ is the internal quality factor of the
uncoupled resonator and $Q_{\rm ext}$ is determined by the value of the
coupling capacitors $C_{\rm ext}$ and the load impedance ($50\,\Omega$ in our
case). By proper choice of $C_{\rm ext}$ the loaded quality factor can be
designed to an optimum value satisfying on the one hand the condition $g \gg
\kappa$  and allowing on the other hand an as high as possible measuring rate
($\propto \kappa$). The loaded quality factor $Q_{\rm L}$ has been extracted
from $S_{21}$ measurements as shown in Fig.~\ref{fig:resonator}(d). In
Fig.~\ref{fig:resonator}(c) we have plotted $Q_{\rm L}$ as a function of the
coupling capacitance $C_{\rm ext}$, calculated using FastCap$^{\circledR}$. The
measured dependence nicely follows the theoretical prediction
\cite{Pozar:2005a} showing that $Q_{\rm L}$ can be deliberately designed over a
wide range. As shown in Fig.~\ref{fig:resonator}(d), internal quality factors
$Q>10^5$ can be achieved with the niobium CPW resonators already at 1.5\,K.

\section{Tunability, Coupling Strength and Symmetry Breaking in Superconducting Circuit QED}
\label{Tunability Coupling Strength Symmetry Breaking}

The field of superconducting circuit QED has attracted considerable attention
over the last years, since it allows to study the interaction of microwave
photons and solid-state quantum circuits on a quantum level. In the following
we consider a classically driven quantum system \cite{Grifoni:1998a} consisting
of a qubit coupled to a harmonic oscillator. We note that the subsequent
discussion is quite general. It is not restricted to the flux qubits discussed
in section~\ref{Fabrication of Superconducting Quantum Circuits}, but is valid
for a general quantum two-level system including both superconducting flux and
charge qubits. The first-order hamiltonian of this system in the diabatic basis
of the qubit eigenstates $\{ |\downarrow\rangle,|\uparrow\rangle \}$ is given
by \cite{Deppe:2008a,Blais:2004a}
\begin{eqnarray}
\widehat{H} & = &  \frac{\epsilon}{2} \widehat{\sigma}_z  - \frac{\Delta_{\rm ge}}{2} \widehat{\sigma}_x
      + \hbar\omega_{\rm r} \left( \hat{a}^\dag \hat{a} + \frac{1}{2} \right)
      \nonumber\\
   &  &
      +\hbar g\,\widehat{\sigma}_z  \left( \hat{a}^\dag + \hat{a} \right)
      +\frac{\Omega}{2} \widehat{\sigma}_z \cos\omega t + \eta \left( \hat{a}^\dag + \hat{a} \right) \cos\omega t
      \,.
   \label{eq:hamiltonianinbarebasis}
\end{eqnarray}
Here, the Hamiltonian $\widehat{H}_{\rm q} \equiv (\epsilon/2)
\widehat{\sigma}_z - (\Delta_{\rm ge}/2) \widehat{\sigma}_x$ describes
the qubit and $\widehat{H}_{\rm r}\equiv \hbar\omega_{\rm r} ( \hat{a}^\dag \hat{a} + 1/2 )$
represents a quantum harmonic oscillator with the photon number
states $|0\rangle, |1\rangle, |2\rangle, \ldots$. $\hat{a}^\dag$ and $\hat{a}$
are the boson creation and annihilation operators and $\widehat{\sigma}_x$ and
$\widehat{\sigma}_z$ the Pauli operators. The level splitting at the degeneracy point $\Delta_{\rm ge} > 0$
by definition. The interaction term between qubit
and resonator is $\widehat{H}_{\rm q,r}=\hbar g\,\widehat{\sigma}_z (
\hat{a}^\dag + \hat{a} )$, and $\widehat{H}_{\rm m,q}=(\Omega/2)
\widehat{\sigma}_z \cos\omega t$ and $\widehat{H}_{\rm m,r}=\eta ( \hat{a}^\dag
+ \hat{a} ) \cos\omega t$ describe the qubit and the resonator microwave
driving, respectively. Furthermore, $\omega_{\rm r}/2\pi$ is the fundamental
resonator frequency, $g/2\pi$ the qubit-resonator coupling strength, and
$\Omega$ and $\eta$ are the qubit and the resonator driving strength. At the
qubit optimal point and, under some approximations, also away from it, the
hamiltonian (\ref{eq:hamiltonianinbarebasis}) is equivalent to that describing
a driven system consisting of an atom coupled to the light field in an optical
cavity. Therefore,  solid-state circuit QED can be considered the circuit
analogue of cavity QED. We note that the hamiltonian
(\ref{eq:hamiltonianinbarebasis}) can be applied to both flux and charge qubits
by choosing an $\epsilon =2I_{\rm p}\delta \Phi_{\rm ex}$ and $\epsilon
=2V_{\rm p}\delta Q_{\rm ex}$, respectively. Here, $V_{\rm p}=e/C_{\rm g}$ is
the voltage associated with an elementary charge on the gate capacitance
$C_{\rm g}$, $\delta \Phi_{\rm ex}$ and $\delta Q_{\rm ex}$ are the applied
gate flux and charge relative to the degeneracy point. The latter is located at
half a flux quantum and half a Cooper pair charge, respectively. In general,
one can switch between the expressions for flux and charge qubits by replacing
the quantities current, flux and inductance by voltage, charge and capacitance.

Most interesting in circuit QED is the so-called strong coupling regime,
$g/2\pi \gg \kappa, \gamma$, where $\kappa$ is the resonator decay rate and
$\gamma$ is the largest of the qubit decay rates. In this situation, the
reversible exchange of microwave photons between a solid-state two-level
quantum system (qubit) and a resonator represents a coherent oscillatory
process. Regarding quantum information processing, this means that quantum
information can be exchanged back and forth many times. To achieve the strong
coupling regime, the qubit-resonator coupling strength $g/2\pi$ must be of the
order of 100\,MHz because presently $\gamma \gtrsim 1\,$MHz, i.e., the decay
rate of the qubit into other channels is still large for superconducting
qubits. As demonstrated in section~\ref{Fabrication of Superconducting Quantum
Circuits}, the fabrication of suitable resonators with $\kappa < 1\,$MHz is
possible. Compared to optical cavity QED, such large values for $g/2\pi$ can be
obtained more easily in superconducting circuit QED for two reasons. First, the
electric and magnetic dipole moments, $\mu_{\rm el}$ and $\mu_{\rm mag}$, of
superconducting charge and flux qubits, respectively, are much larger than
those of natural atoms. Second, the mode volume $V_{\rm m}$ of transmission
line microwave resonators (cf. Fig.~\ref{fig:resonator}) can be made much
smaller than for 3D cavities in quantum optics, resulting in large electric or
magnetic fields. Note that the fields associated with vacuum fluctuations are
$E_0 = \sqrt{\hbar\omega_{\rm r} /2\epsilon_0V_{\rm m}}$ and
$B_0 = \sqrt{\mu_0\hbar\omega_{\rm r} /2V_{\rm m}}$. Since mode volumes as small as
$10^{-12}\textrm{m}^3$ seem to be feasible, the electric and magnetic fields
associated with vacuum fluctuations are of the order of 0.1\,V/m and 1\,nT for
$\omega_{\rm r} /2\pi$ of a few GHz.

In the following we first address the tunability of the energy levels of
superconducting charge and flux qubits by static electric and magnetic fields.
The derived energy shifts are equivalent to the dc Stark and Zeeman shifts in
natural atoms. Then, we discuss the atom-resonator coupling strength and some
fundamental symmetry aspects of superconducting circuit QED. In our discussion
we use a semiclassical approach, treating the microwave driving as classical
fields.

\subsection{Tunability of Charge and Flux Qubits}

In this section we compare artificial superconducting atoms (charge and flux
qubits) to natural atoms (we use hydrogen for simplicity) regarding their
tunability by static fields. In Table~\ref{tab:energyshift} we have summarized
the first- and second-order energy shifts $\Delta E^{(1)}_{\rm el,mag}$ and
$\Delta E^{(2)}_{\rm el,mag}$ due to electric and magnetic fields, $\delta
E_{\rm ex}$ and $\delta B_{\rm ex}$, applied via suitable control gates.
Discussing the first-order terms, we consider the charge qubit at the
degeneracy point as a system of two degenerate levels $|\downarrow\rangle$ and
$|\uparrow\rangle$ corresponding to adjacent Cooper-pair number states with
Cooper pair numbers $N$ and $N+1$. At this point, the polarization charge
$Q_{\rm ex} = 2eN_{\rm ex}$ induced by an applied gate voltage corresponds to
$2e(N+1/2)$ for both charge basis states. We define $\delta N_{\rm ex}
= N_{\rm ex}-(N+1/2)$ as the deviation from the Cooper pair number at
the degeneracy point. Therefore, at the degeneracy point we have $\delta N_{\rm
ex}=0$ and, correspondingly, $\delta E_{\rm ex}=0$, where $\delta E_{\rm ex} =
\delta Q_{\rm ex}/C_{\rm g}d = 2e\delta N_{\rm ex}/C_{\rm g}d$ is the electric
field applied via a control gate, which just shifts the Cooper pair number on
the island by $\delta N_{\rm ex}$ away from the degeneracy point. Here, $C_{\rm
g}$ is the gate capacitor and $d$ the distance between the control gate and the
island. Our discussion shows that we can consider the charge qubit at the
degeneracy point as a degenerate two-level system in a zero effective control
electric field $\delta E_{\rm ex} =0$. The finite coupling of the charge states
$|\downarrow\rangle$ and $|\uparrow\rangle$ results in the symmetric and
antisymmetric superposition states $|\textrm{g}\rangle = (|\downarrow\rangle +|\uparrow\rangle)/\sqrt{2}$ and $|\textrm{e}\rangle=(|\downarrow\rangle-|\uparrow\rangle)/\sqrt{2}$, respectively, with an energy splitting
$\Delta_{\rm ge}$ of their eigenenergies. The situation for the flux qubit is
completely analogous. At the degeneracy point it can be considered as a system
of two degenerate levels $|\downarrow\rangle$ and $|\uparrow\rangle$
corresponding to the persistent currents $\pm I_{\rm p}$ circulating in the
qubit loop. The polarization flux $\Phi_{\rm ex} = N_{\rm ex} \Phi_0$ is
generated by a magnetic field $B_{\rm ex}$ applied via a control gate. Again,
at the degeneracy point we have $\delta N_{\rm ex}=0$ and $\delta B_{\rm
ex}=0$. Here, $\delta B_{\rm ex} = \delta \Phi_{\rm ex}/A = \delta N_{\rm
ex}\Phi_0/A$ is the magnetic control field, which just changes the flux number
in the loop of area $A$ by $\delta N_{\rm ex}$. Evidently, at the degeneracy
point we can consider the flux qubit as a two-level system in a zero effective
magnetic control field $\delta B_{\rm ex} =0$. As for the charge qubit, the
finite coupling of the flux states results in superposition states with
eigenenergies separated by $\Delta_{\rm ge}$.

\begin{table}
\caption{\label{tab:energyshift} Characteristic first- and second-order shifts
of the energy levels due to external electric and magnetic fields $\delta
E_{\rm ex}$ and $\delta B_{\rm ex}$ for natural atoms (hydrogen) compared to
those of superconducting charge and flux qubits. For the estimates of the
electric and magnetic dipole moments $\mu_{\rm el}$ and $\mu_{\rm mag}$ as well
as the polarizabilities $\chi_{\rm el}$ and $\chi_{\rm mag}$ we used the
typical values $d=0.1-1\,\mu$m for the distance between island and gate
electrode (charge qubit), as well as $A=5-10\,\mu\textrm{m}^2$ and $I_{\rm p} =
100 -500\,$nA for the loop area and the persistent current (flux qubit),
respectively. Furthermore, $\mu_0=1.26\times 10^{-6}$Vs/Am is the vacuum
permeability, $\epsilon_0=8.85\times 10^{-12}$As/Vm the vacuum permittivity,
$a_{\rm B}=5.29\times 10^{-11}$m the Bohr radius, $m=9.109\times 10^{-31}$kg
the electron mass, and $e=1.602\times 10^{-19}$As the electron charge. }

\lineup
\begin{tabular}{{l} {c} {c} {c} }
 \br
energy & natural atom & charge qubit & flux qubit
 \\
 shift &  &  &
 \\ \br
     $\Delta E_{\rm el}^{(1)}$
   & $\pm \mu_{\rm el} \cdot \delta E_{\rm ex}$
   & $\pm \mu_{\rm el} \cdot \delta E_{\rm ex}$
   & ---
 \\   \bs
   & $\langle\mu_{\rm el}\rangle =0$
   & $\mu_{\rm el}  = e d$
   &
 \\    \bs
   &
   & $\sim 10^{-26}$-$10^{-25}$Cm
   &
\\ \mr
     $\Delta E_{\rm mag}^{(1)}$
   & $\pm \mu_{\rm mag} \cdot \delta B_{\rm ex}$
   & ---
   & $\pm \mu_{\rm mag} \cdot \delta B_{\rm ex}$
 \\ \bs
   & $\mu_{\rm mag} \simeq \mu_{\rm B}$
   &
   & $\mu_{\rm mag} = I_{\rm p} A$
 \\  \bs
   &
   &
   & $\sim 10^{4}$-$10^{5}\mu_{\rm B}$
\\ \mr
     $\Delta E_{\rm el}^{(2)}$
   & $- \frac{1}{2} \chi_{\rm el} \cdot \delta E_{\rm ex}^2$
   & $- \frac{1}{2} \chi_{\rm el} \cdot \delta E_{\rm ex}^2$
   & ---
\\ \bs
   & $\chi_{\rm el} \simeq 4\pi \epsilon_0 a_{\rm B}^3 $
   & $\chi_{\rm el} \simeq -2 \case{(ed)^2}{\Delta_{\rm ge}}$
   &
\\ \bs
   & $\sim 10^{-41}\textrm{Cm}^2/\textrm{V}$
   & $\sim 10^{-28}$-$10^{-26}$Cm$^2$/V
   &
\\ \mr
     $\Delta E_{\rm mag}^{(2)}$
   & $- \frac{1}{2} \chi_{\rm mag} \cdot \delta B_{\rm ex}^2$
   & ---
   & $- \frac{1}{2} \chi_{\rm mag} \cdot \delta B_{\rm ex}^2$
\\ \bs
   & $\chi_{\rm mag} \simeq - \case{e^2 a_{\rm B}^2}{6m}$
   &
   & $\chi_{\rm mag} \simeq - 2 \case{(I_{\rm p}A)^2}{\Delta_{\rm ge}}$
\\ \bs
   & $\sim 10^{-29}$Am$^2$/T
   &
   & $\sim 10^{-13}$-$10^{-11}$Am$^2$/T
\\ \br
\end{tabular}
\end{table}

We now can discuss the interaction energy of the charge and flux qubit with
static control fields $\delta F_{\rm ex} = \{\delta E_{\rm ex}, \delta B_{\rm
ex}\}$. Since both the electric and magnetic dipole moments in the two basis
states $|\downarrow\rangle$ and $|\uparrow\rangle$ have opposite sign, the
interaction operator in the basis $\{ |\downarrow\rangle ,|\uparrow\rangle \}$
can be written as
\begin{equation}
\widehat{W} = - \mu \, \delta F_{\rm ex} \; \widehat{\sigma}_z
\label{eq:dipoleoperator}
\end{equation}
with the dipole moments $\mu = \{\mu_{\rm el} , \mu_{\rm mag}\}$ given by
\begin{eqnarray}
\mu _{\rm el} & = & e \, d
\label{eq:electricdipolemoment}
\\
\mu _{\rm mag} & = & I_{\rm p} \, A \;\; .
\label{eq:magneticdipolemoment}
\end{eqnarray}
We immediately see that the first-order energy shifts of the uncoupled levels
$|\downarrow\rangle$ and $|\uparrow\rangle$ are given by the products
\begin{eqnarray}
\Delta E_{\rm el}^{(1)} & = & \pm \mu_{\rm el} \cdot \delta E_{\rm ex}
\\
\Delta E_{\rm mag}^{(1)} & = & \pm \mu_{\rm mag} \cdot \delta B_{\rm ex} \;\; .
\end{eqnarray}
For natural atoms, the corresponding energy shifts are the linear Stark and
Zeeman shifts. Note that natural atoms and molecules with an inversion center
in a non-degenerate electronic state do not have a permanent electric dipole
moment and therefore have $\Delta E^{(1)}_{\rm el} =0$, that is, they do not
show a linear Stark effect. As indicated by the numbers listed in
Table~\ref{tab:energyshift}, the dipole moments of the qubits are by many
orders of magnitudes larger than those of natural atoms. For the charge qubit,
$\mu_{\rm el} = e d$ is determined by the distance $d$ between the island and
the gate electrode and can be as large as $10^4$ to $10^5$ debye
($1\,\textrm{debye} = 3.3\times 10^{-30}$Cm). In other words, it exceeds the
typical values for natural atoms by 4 to 5 orders of magnitude. The same is
true for the magnetic dipole moment, which is given by the persistent current
$I_{\rm p}$ circulating around the qubit loop and the loop area $A$. It
typically amounts to $10^4$ to $10^5$ Bohr's magnetons compared to only a
single Bohr's magneton for natural atoms. Obviously, the large dipole moments
of the qubits are caused by their much larger physical size. Note that $d$ and
$A$ are of the order of $\mu$m and $\mu$m$^2$ for the qubits, but only {\AA} and
{\AA}$^2$ for natural atoms. Of course, the large dipole moments of qubits also
have a disadvantage. They make the qubits more sensitive to electric and
magnetic noise. We also note that large dipole moments are obtained for Rydberg
atoms \cite{Meschede:1985a,Rempe:1987a,Raimond:2001a}, which are highly excited
atomic states with much larger atomic diameter.

So far we have considered only the uncoupled basis states. The finite coupling
of $|\downarrow\rangle$ and $|\uparrow\rangle$ results in the superposition
states \cite{Cohen-Tannoudji:1977}
\begin{eqnarray}
|g\rangle & = & +\cos \frac{\theta}{2}\textrm{e}^{+i\varphi /2} |\downarrow\rangle
                -\sin \frac{\theta}{2} \textrm{e}^{-i\varphi /2} |\uparrow\rangle
\\
 |e\rangle & = & + \sin \frac{\theta}{2}\textrm{e}^{+i\varphi /2}|\downarrow\rangle
                 + \cos \frac{\theta}{2}\textrm{e}^{-i\varphi /2} |\uparrow\rangle
\end{eqnarray}
with the level spacing
\begin{equation}
E_{\rm ge} = \sqrt{\epsilon^2 + \Delta_{\rm ge}^2} \; .
\label{eq:levelsplitting}
\end{equation}
Here, $\tan \theta \equiv  \Delta_{\rm ge}/\epsilon$ and $\epsilon \equiv
2\mu\; \delta F_{\rm ex}$ is the energy bias given by
\begin{eqnarray}
\epsilon_{\rm el} & \equiv & 2ed \; \delta E_{\rm ex} = \frac{(2e)^2}{2C_{\rm g}} \; 2\delta N_{\rm ex}
\\
\epsilon_{\rm mag} & \equiv & 2I_{\rm p}A \; \delta B_{\rm ex} = \frac{\Phi_0^2}{2L} \; 2\delta N_{\rm ex}
\end{eqnarray}
for the charge and flux qubit, respectively. The large dipole moments of the
qubits allow to considerably tune $E_{\rm ge}$ by applied control fields.
Already fields of the order of $10^3$V/m and $10^{-4}$T are sufficient to cause
changes of the order of $\Delta_{\rm ge}$, whereas for natural atoms the
equivalent shifts are usually negligibly small. It is important to note that
the expectation values are $\langle g| \widehat{\sigma}_z|g\rangle = -\cos\theta$
and $\langle e| \widehat{\sigma}_z|e\rangle = \cos\theta$. Since $\cos\theta
=0$ at the degeneracy point, the expectation value of the dipole moments vanish
for the superposition states. Therefore, the energy shifts $\Delta E_{\rm
el}^{(1)}$ and $\Delta E_{\rm mag}^{(1)}$ disappear, making the charge and flux
qubit insensitive to low-frequency fluctuations (e.g. $1/f$ noise) of the
electric and magnetic field at this operation point,
respectively~\cite{Kakuyanagi:2007a,Deppe:2007a,Yoshihara:2006a}. Nevertheless,
there are significant second-order contributions left, originating from the
large polarizabilities of the qubits as discussed in the following.

The second-order terms can be expressed by using the polarizabilities of the
natural and artificial atoms. For natural atoms with spherical symmetry, the
polarizability tensor is isotropic, giving the quadratic terms $\Delta E_{\rm
el}^{(2)} = - \frac{1}{2} \chi_{\rm el} \cdot \delta E_{\rm ex}^2$ and $\Delta
E_{\rm mag}^{(2)} = - \frac{1}{2} \chi_{\rm mag} \cdot \delta B_{\rm ex}^2$,
where $\chi_{\rm el}$ and $\chi_{\rm mag}$ are the electric and magnetic
polarizabilities, respectively. Using second-order perturbation theory, the
polarizabilities of the qubits can be written as
\begin{equation}
\chi_{\rm el,mag} = -2 \frac{\langle g|\widehat{\mu}_{\rm el,mag}|e\rangle
                          \langle e|\widehat{\mu}_{\rm el,mag}|g\rangle}{E_{\rm ge}}
\end{equation}
with $E_{\rm ge} \equiv E_{\rm e} - E_{\rm g}$, $\widehat{\mu}_{\rm el} = ed
\widehat{\sigma}_z$, and $\widehat{\mu}_{\rm mag} = I_{\rm p}A
\widehat{\sigma}_z$. Because $\langle g| \widehat{\sigma}_z|e\rangle = \langle
e| \widehat{\sigma}_z|g\rangle = \sin\theta$, we obtain the polarizabilities
\begin{eqnarray}
\chi_{\rm el} & = & - 2 \frac{(ed)^2 \; \sin^2\theta}{E_{\rm ge} }
\\
\chi_{\rm mag} & = & - 2 \frac{(I_{\rm p}A)^2 \; \sin^2\theta}{E_{\rm ge} }
\end{eqnarray}
and the second-order energy shifts
\begin{eqnarray}
\Delta E_{\rm el}^{(2)} & = & \frac{(ed)^2 \; \sin^2\theta}{E_{\rm ge}} \cdot \delta E_{\rm ex}^2
\\
\Delta E_{\rm mag}^{(2)} & = & \frac{(I_{\rm p}A)^2 \; \sin^2\theta}{E_{\rm ge}} \cdot \delta B_{\rm ex}^2 \;\; .
\end{eqnarray}
Interestingly, the second-order contributions are maximum at the degeneracy
point, where the first-order contributions vanish,  and decrease moving away
from it because of the $\sin^2\theta$ term. Close to the degeneracy point we
can use the approximations $\sin\theta \simeq 1$ and $E_{\rm ge} \simeq
\Delta_{\rm ge}$ and obtain the polarizabilities $\chi_{\rm el} \simeq -
2(ed)^2/\Delta_{\rm ge}$ and $\chi_{\rm mag} \simeq -2(I_{\rm
p}A)^2/\Delta_{\rm ge}$, which are listed in Table~\ref{tab:energyshift}.
Again, due to the much larger size of the qubits the polarizabilities are by
many orders of magnitudes larger than those of natural atoms. Since in
electrostatics the electric and magnetic energy contributions also can be
expressed as $\frac{1}{2} C_{\rm Q}d^2 \delta E_{\rm ex}^2$ and $\frac{1}{2}
(A^2/L_{\rm Q})\delta B_{\rm ex}^2$, the polarizabilities can be associated
with a quantum capacitance $C_{\rm Q} = 2e^2/\Delta_{\rm ge}$ and a quantum
inductance $L_{\rm Q} = \Delta_{\rm ge}/2I_{\rm p}^2$, respectively
\cite{Averin:2003a,Sillanpaa:2005a,Duty:2005a,Koenemann:2007a}.

\subsection{Coupling Strength and Symmetry Breaking in Circuit QED}

In the previous subsection we have seen that charge and flux qubits show a
large tunability of their energy levels due to their large dipole moments and
polarizabilities, allowing to tune the level separation over a wide range by
\emph{static} electric and magnetic gate fields. In the following we address
the coupling strength with \emph{dynamic electric and magnetic fields} as well
as some fundamental symmetry aspects of superconducting circuit QED. In our
analysis we do not only focus on flux qubits discussed in section~\ref{Flux
Qubits}. We rather show that the arguments given in the following are quite
general and can be straightforwardly applied also to charge qubits.

\begin{figure}[tb]
 \centering{
    \includegraphics[width=0.8\columnwidth]{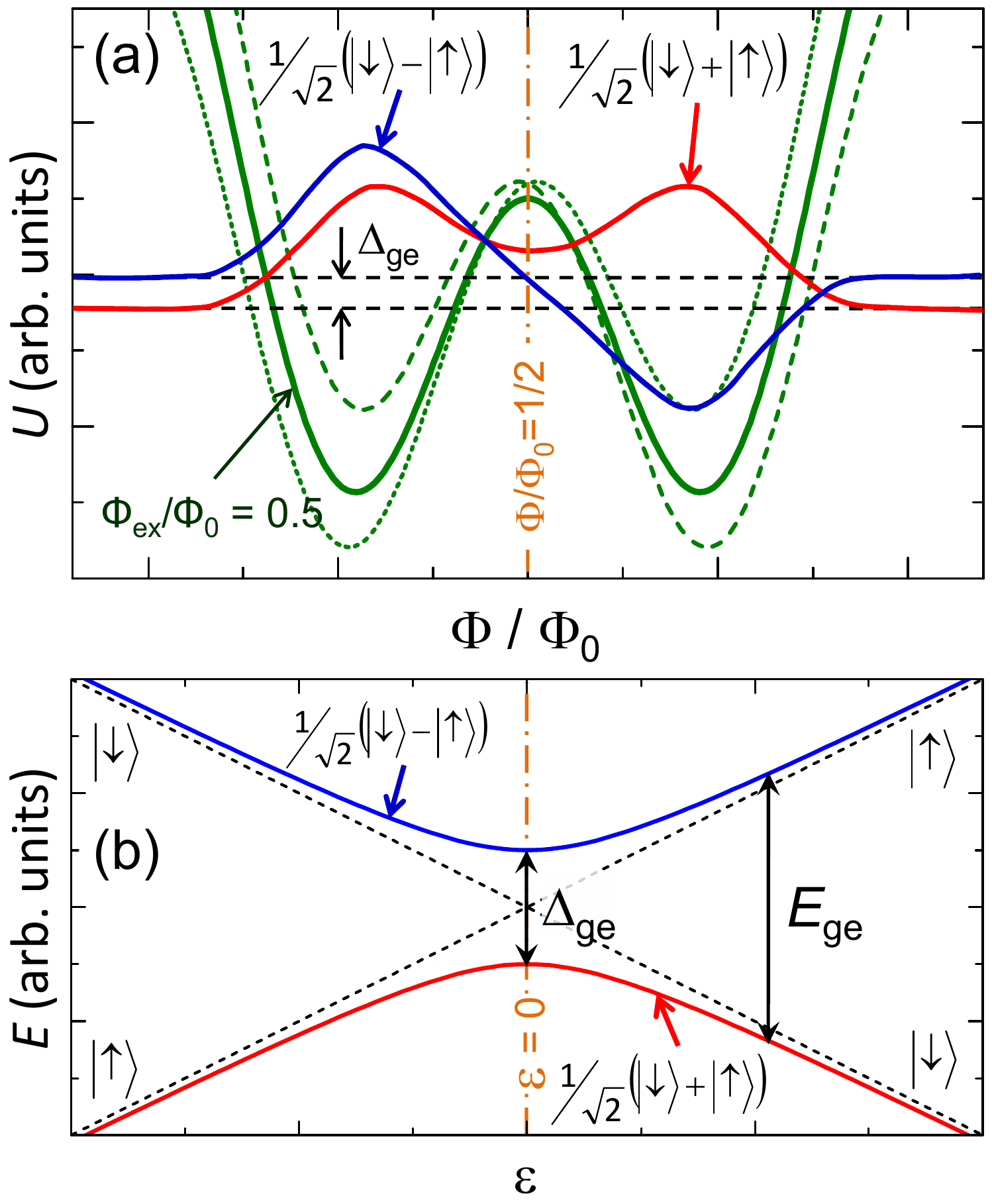}}
    \caption{(a) Sketch of the double-well potential $U$ of a flux qubit plotted versus the total flux $\Phi$ in the qubit loop
             for an externally applied flux $\Phi_{\rm ex}=\Phi_0/2$ (solid green line). Also shown are the two eigenfunctions
             of the ground state (symmetric, red) and the first excited state (antisymmetric, blue).
             The dotted and dashed lines show the qubit potential for an external flux larger and smaller than $\Phi_0/2$.
             (b) The energy $E$ of the two superposition states $|\textrm{g}\rangle =1/\sqrt{2} (|\downarrow\rangle +|\uparrow\rangle)$
             and $|\textrm{e}\rangle=1/\sqrt{2} (|\downarrow\rangle -|\uparrow\rangle)$ versus the energy bias $\epsilon$. The dashed lines
             show the energies of the uncoupled basis states.
             }
    \label{fig:potential}
\end{figure}

It is well known \cite{Orlando:1999a} that the potential of the three-junction
flux qubit can be reduced to the one-dimensional double-well $U(\Phi)$ shown in
Fig.~\ref{fig:potential}(a). Here, $\Phi$ is the total magnetic flux in the
qubit loop. It originates from the externally applied flux $\Phi_{\rm ex}$ and
the flux due to the current circulating in the loop, having an expectation
value $I_p \langle \widehat{\sigma}_z \rangle$. We first discuss the situation
at the degeneracy point, where the external flux $\Phi_{\rm ex}$ equals
$\Phi^{(n)} \equiv \Phi_0(n+1/2)$. Here, $n$ is an integer and we only consider
$n=0$ in the following. Then, the two wells are symmetrical with respect to
$\Phi = \Phi_0/2$ and $I_p \langle \widehat{\sigma}_z \rangle$ vanishes. In
Fig.~\ref{fig:potential}(a) we also plot the eigenfunctions
$|\textrm{g}\rangle$ and $|\textrm{e}\rangle$, which, at the degeneracy point,
are fully symmetric and antisymmetric superpositions of the two basis states
$|\downarrow\rangle$ and $|\uparrow\rangle$ with clock- and counterclockwise
circulating persistent currents $\pm I_{\rm p}$. For a wider region around the
degeneracy point, the energy difference $E_{\rm ge}(\epsilon)$ (cf.
eq.(\ref{eq:levelsplitting})) between the two levels is shown in
Fig.~\ref{fig:potential}(b). At the degeneracy point, the energy bias $\epsilon
= 2I_{\rm p}A \delta B_{\rm ex} =2I_{\rm p} \delta\Phi_{\rm ex}$ is zero
resulting in $E_{\rm ge}=\Delta_{\rm ge}$. The probability to observe
$|\downarrow\rangle$ ($|\uparrow\rangle$) is just 0.5 at the degeneracy point
and approaches 1 or 0 (0 or 1) when decreasing or increasing the applied flux.
For the charge qubit the situation is completely analogous. At the degeneracy
point, the ground state $|\textrm{g}\rangle$ and the excited state
$|\textrm{e}\rangle$ are fully symmetric and antisymmetric superpositions of
the two basis states $|\downarrow\rangle$ and $|\uparrow\rangle$ corresponding
to neighboring Cooper pair number states on the superconducting island.

Considering symmetries, we immediately see from Fig.~\ref{fig:potential}(a)
that the potential $U$ of the flux qubit is fully symmetric with respect to the
variable $\delta\Phi = \Phi - \Phi_0/2$ for an external flux bias $\Phi_{\rm
ex}= \Phi_0/2$. That is, at the degeneracy point we have
$U(\delta\Phi)=U(-\delta\Phi)$. We note that $\delta\Phi$ is associated with
the phase differences $\gamma_1$ and $\gamma_2$ across the two larger junctions
as $(\gamma_2-\gamma_1)/2 = 2\pi (\delta\Phi/\Phi_0)$. In complete analogy,
$U(\delta Q)=U(-\delta Q)$ for the charge qubit at an external charge bias of
$Q_{\rm ex} = 2e/2$. Here, $\delta Q = Q-2e/2$ is the quantum variable
conjugate to $\delta\Phi$. In both cases the qubit eigenstates
$|\textrm{g}\rangle$ and $|\textrm{e}\rangle$ have even and odd parity because
they are fully symmetric and antisymmetric superpositions of the basis states.
Then we can write
\begin{eqnarray}
\widehat{\Pi} \; |\textrm{g}\rangle & = & + |\textrm{g}\rangle
\\
\widehat{\Pi} \; |\textrm{e}\rangle & = & - |\textrm{e}\rangle \; ,
\label{eq:parityoperator}
\end{eqnarray}
where $\widehat{\Pi}$ is the parity operator. Moving away from the degeneracy
point, the potential is tilted to the left or right as indicated by the dotted
and dashed lines in Fig.~\ref{fig:potential}(a). Then, symmetry is no longer
well defined and $|\textrm{g}\rangle$ and $|\textrm{e}\rangle$ are no longer
fully symmetric and antisymmetric superpositions of the basis states. In
this way, we we can break the symmetry of the flux qubit in a controlled way
using an external control parameter, namely the static magnetic flux
$\delta\Phi_{\rm ex}=\Phi_{\rm ex}-\Phi_0/2$ generated by a magnetic field
applied via a control gate. For charge qubits, correspondingly the control
parameter is the excess charge $\delta Q_{\rm ex}=2e\delta N_{\rm ex}=
2e(N_{\rm ex}-1/2)$ induced on the island by an electric field applied via a
gate capacitor.

The controlled symmetry breaking by static applied gate fields has far-reaching
consequences on the selection rules for level transitions. At the degeneracy
point,  these selection rules can be intuitively understood by the fact that
the qubit potential exhibits mirror symmetry, $U(\delta\Phi)=U(-\delta\Phi)$
and $U(\delta Q)=U(-\delta Q)$ for the flux and charge qubit, respectively [cf.
Fig.~\ref{fig:potential}(a)]. Hence, the interaction operator of the one-photon
driving initiating level transitions between the fully symmetric and
antisymmetric states $|g\rangle$ and $|e\rangle$ must be odd with respect to
the variable $\delta\Phi_{\rm ex}$ or $\delta Q_{\rm ex}$
\cite{Deppe:2008a,Liu:2005a}. Such odd-parity interaction operators are the
dipole operators [cf. eqs.(\ref{eq:dipoleoperator}),
(\ref{eq:electricdipolemoment}), and (\ref{eq:magneticdipolemoment})] for the
flux and charge qubit, respectively, because $\sum_{g,e} |i\rangle\langle i|
\widehat{\mu}_{\rm mag, el} |j\rangle\langle j| \propto \widehat{\sigma}_x$
and, hence, the anticommutator $\{\widehat{\Pi},\widehat{\mu}_{\rm mag, el}\} =
0$. In contrast the interaction operator for the two-photon driving is even
\cite{Deppe:2008a}. Hence, if the interaction operator is a dipole operator, at
the degeneracy point only one-photon transitions are allowed, whereas
two-photon transitions are strictly forbidden. These selection rules correspond
to those for electric dipole transitions in natural atoms
\cite{Cohen-Tannoudji:1977}. Away from the degeneracy point, the qubit
potential no longer exhibits mirror symmetry with respect to $\delta\Phi$ and
$\delta Q$ for the flux and charge qubit, respectively, and the superposition
states $|g\rangle$ and $|e\rangle$ no longer are purely symmetric and
antisymmetric states. Then, the strict selection rules valid at the degeneracy
point do no longer apply and one- and two-photon transitions can coexist.

To gain a better insight into this issue one has to explicitly discuss the
matrix elements for level transitions in a flux qubit induced by an external
microwave driving $\delta B_{\rm ex}(t) = \delta B_0 \cos\omega t$ and $\delta
E_{\rm ex}(t) = \delta E_0 \cos\omega t$ of the flux and charge qubit,
respectively. In this context, we first note that for the flux qubit the
transitions induced by $\delta B_{\rm ex}(t)$ are of \emph{electric} dipole
type. This can be understood by recalling that for the flux qubit the magnetic
flux $\Phi$ is equivalent to the spatial coordinate, whereas $Q$ is equivalent
to the momentum. In contrast, for the charge qubit, the charge $Q$ is
equivalent to the spatial coordinate and the flux $\Phi$ to the momentum. With
the dipole operator $\widehat{D} (t)= \frac{\Omega}{2} \cos\omega t
\;\widehat{\sigma}_z$ and using $\langle g| \widehat{\sigma}_z|e\rangle =
\sin\theta$ we obtain the matrix element for one-photon transitions
\cite{Cohen-Tannoudji:1977}
\begin{eqnarray}
\langle g|\widehat{D} (t)|e \rangle &=&  \frac{\Omega}{2}
\; \langle g|\widehat{\sigma}_z|e\rangle \; \,\cos\omega t
\nonumber \\
& = &  \frac{\Omega}{2} \,\sin\theta \; \cos\omega t \; ,
\label{eq:matrixelement1}
\end{eqnarray}
where we define $\Omega =2I_{\rm p}A \delta B_0$ and $\Omega =2V_{\rm p}C d
\delta E_0 = 2ed\delta E_0$ for the flux and charge qubit, respectively. Hence,
the modulus of the matrix element for one-photon transitions is given by
\begin{equation}
D^{(1)} = \frac{\Omega}{2} \; \sin\theta \; .
\label{eq:matrixelement2}
\end{equation}
For both the flux and charge qubit the transition matrix element is
proportional to $\sin\theta = \Delta_{\rm ge}/E_{\rm ge}$. Hence, it is maximum
at the degeneracy point ($\sin\theta =1$) and continuously decreases when
moving away from it.

The treatment of the two-photon transitions is more complicated. For $2\omega
\simeq \omega_{\rm ge}=E_{\rm ge}/\hbar$ (resonant two-photon driving) the
level transitions can be described as a two-step process and a detuning of $E_{\rm
ge}/2$ \cite{Brune:1987a}. Then, the matrix element for the two-photon
transition is obtained to
\begin{eqnarray}
\frac{\langle g|\widehat{D}(t)|e \rangle
                        \langle e|\widehat{D}(t)|e \rangle }{E_{\rm ge}/2}
 & = &
\frac{\Omega^2}{4} \; \frac{\langle g|\widehat{\sigma}_z|e \rangle
                        \langle e|\widehat{\sigma}_z|e \rangle }{E_{\rm ge}/2} \; \,\cos^2\omega t \;\; .
\label{eq:matrixelement3}
\end{eqnarray}
With $\langle g|\widehat{\sigma}_z|e \rangle = \sin\theta$, $\langle
e|\widehat{\sigma}_z|e \rangle = \cos\theta$, $\cos^2 \omega t =\frac{1}{2}(1
+\cos 2\omega t)$ and $E_{\rm ge} = \Delta_{\rm ge} /\sin\theta$ we obtain
\begin{eqnarray}
\frac{\langle g|\widehat{D}(t)|e \rangle \langle e|\widehat{D}(t)|e \rangle }{E_{\rm ge}/2}
 & = &
 \frac{\Omega^2}{4\Delta_{\rm ge}} \; \sin^2\theta \cos\theta (1+\cos 2\omega t) \;\; .
\label{eq:matrixelement4}
\end{eqnarray}
From this expression, we identity the modulus of the matrix element for
two-photon transitions
\begin{equation}
D^{(2)} = \frac{\Omega^2}{4\Delta_{\rm ge}} \; \sin^2\theta\cos\theta \; .
\label{eq:matrixelement5}
\end{equation}
The transition matrix elements for both flux and charge qubits are now
proportional to $\sin^2\theta\cos\theta =\Delta^2_{\rm ge} \epsilon /E^3_{\rm
ge}$. Because $\epsilon =0$ at the degeneracy point, no two-photon transitions
are allowed there in agreement with the qualitative discussion above. However,
moving away from the degeneracy point the matrix element $D^{(2)}$ becomes
finite, making a two-photon driving of the qubit possible. This has been
demonstrated in a recent experiment \cite{Deppe:2008a}, where a flux qubit
coupled to an $LC$ resonator has been studied.

\begin{figure}[tb]
 \centering{
    \includegraphics[width=0.9\columnwidth]{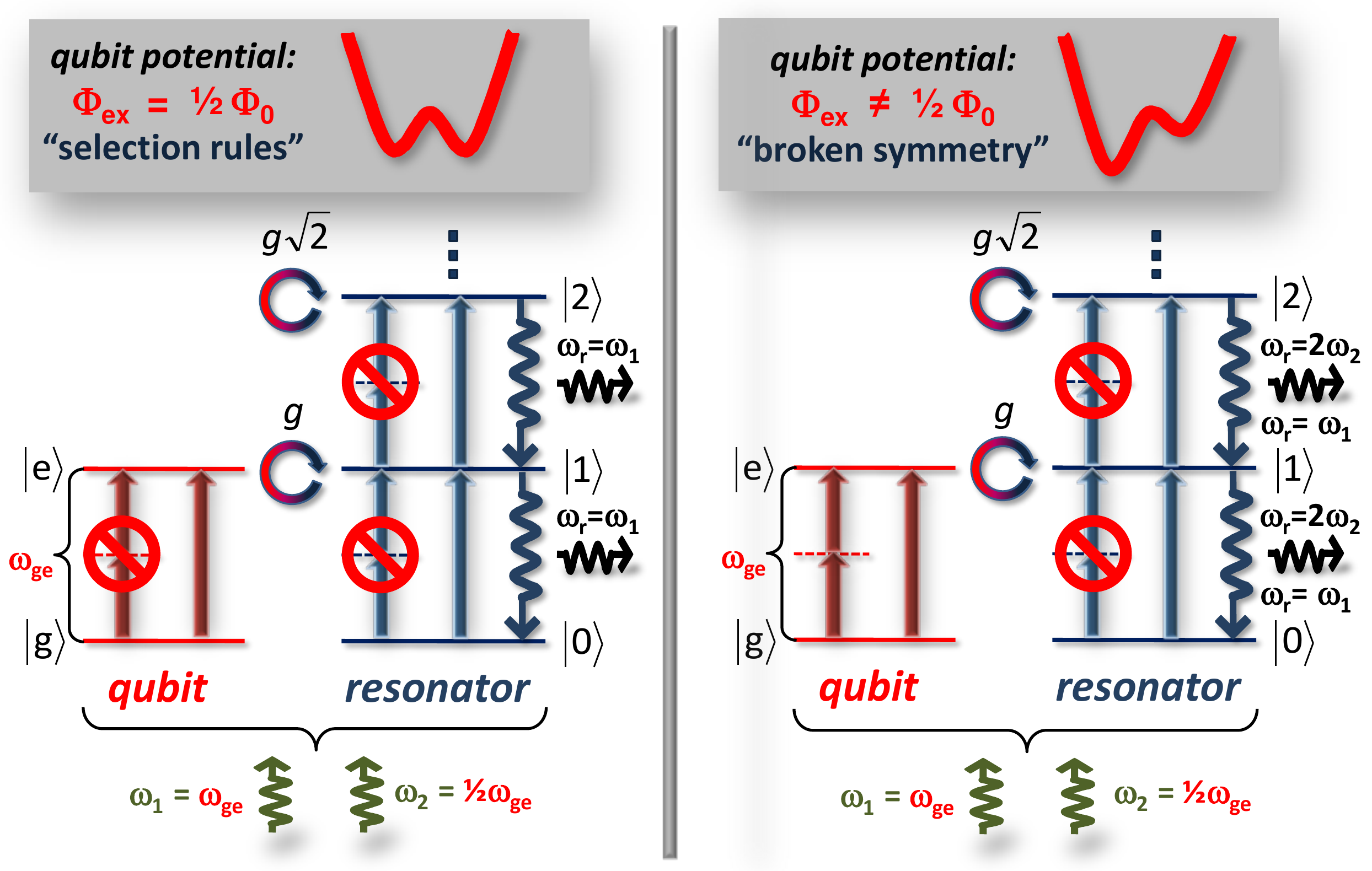}}
    \caption{Selection rules for one- and two-photon transitions in a flux qubit having a potential
             with (left) and without (right) mirror symmetry.
             In the symmetric situation (left) the two-photon process is forbidden. Since
             two-photon transitions are also forbidden for the harmonic oscillator,
             qubit and resonator can be excited only resonantly at frequency $\omega_1=\omega_{\rm ge}=\omega_{\rm r}$.
             In contrast, in the case of broken symmetry (right) the qubit can be excited
             both by one- and two-photon processes. Therefore, also for $\omega_2=\omega_{\rm ge}/2$
             (two-photon process) the excited qubit state $|e\rangle$ can be populated.
             Due to the qubit-resonator coupling $g/2\pi$ a state transfer to the resonator is achieved.
             The resonator state, in turn, decays by emission of a photon of twice the frequency
             (frequency up-conversion).
             }
    \label{fig:selectionrules}
\end{figure}

The possibility to deliberately tune the level spacing and the symmetry
properties of flux and charge qubits by an external control field can be used
to study the effect of symmetry breaking in superconducting circuit QED
systems. The key effect of symmetry breaking on selection rules and matrix
elements for one- and two-photon transitions in a coupled flux qubit-resonator
system is sketched in Fig.~\ref{fig:selectionrules} for a three-junction flux
qubit. The resonator, a harmonic oscillator, can never accept a two-photon
driving. Furthermore, for an external flux bias $\Phi_{\rm ex} = \Phi_0/2$ the
qubit potential has perfect mirror symmetry (left) and the selection rules
discussed above hold. Then, qubit and resonator can be excited only resonantly
at frequency $\omega_1 = \omega_{\rm ge}$. By changing the flux bias away form
$\Phi_0/2$, the mirror symmetry of the potential is broken (right). In this
case, also for two-photon driving ($\omega_2=\omega_{\rm ge}/2$) the excited
qubit state $|e\rangle$ can be populated. The strong qubit-resonator coupling
$g/2\pi$ results in a state transfer to the resonator, which subsequently
decays by emission of a photon of twice the frequency. This frequency
up-conversion mechanism for microwave photons in a coupled qubit-resonator
system has been studied in a recent experiment \cite{Deppe:2008a}. We note that
the occurrence of this process is a direct consequence of the underlying
symmetry properties of the system, which can be chosen on purpose by an
external control parameter (magnetic flux in our case). The symmetry properties
of the qubit-resonator system have an interesting consequence. Depending on the
choice of $\Phi_{\rm ex}$ and the microwave frequency $\omega/2\pi$, one can
choose to drive either only the qubit, only the resonator, or both the qubit
and the resonator. This selective driving once more demonstrates the benefits
of the tunability of solid-state circuit QED systems. Besides its role in
fundamental research, the controlled symmetry breaking in artificial quantum
systems has good prospects of use in numerous applications. Particular examples
are parametric up-conversion, the generation of single microwave photons on
demand \cite{Mariantoni:2005a,Houck:2007a,Liu:2004a}, and squeezing of quantum
states \cite{Moon:2005a}.

\section{Conclusions}

We have fabricated superconducting three-junction flux qubits using electron
beam lithography and two-angle shadow evaporation. The qubits are based on
Al/AlO$_x$/Al Josephson junctions with areas down to $0.02\,\mu\textrm{m}^2$
and high current densities above 1\,kA/cm$^2$. They show a level splitting above
4\,GHz at 50\,mK. We also fabricated readout dc SQUIDs and additional on-chip
circuit elements such as shunting capacitors for the readout SQUID,
transmission lines for the application of microwave signals to the qubits, and
coplanar waveguide resonators. The latter are based on Nb films and have
resonance frequencies ranging between about 1 and 10\,GHz. They show internal
quality factors above $10^5$ at 1.5\,K.

We also studied the tunability of superconducting charge and flux qubits by
external control fields. We derived the effective dipole moments of the qubits
and show that they are by several orders of magnitude larger than those of
natural atoms because of the much larger geometrical size of the artificial
solid-state atoms (qubits). Furthermore, we derived the first- and second-order
energy shifts of the qubit levels due to control fields as well as expressions
for the matrix elements describing the level transitions at and away from the
degeneracy point. We finally addressed some fundamental symmetry aspects of
superconducting qubits. We argue that at the degeneracy point the qubit
potential has mirror symmetry resulting in selection rules for level
transitions equivalent to those in natural atoms. However, the symmetry of the
qubit potential can be easily broken by external control fields. Then, the
strict selection rules are no longer valid. In particular, two-photon
transitions are allowed resulting in an interesting up-conversion dynamics in a
coupled qubit-resonator system, which we could observe in a recent experiment.

\ack

This work is supported by the German Science Foundation via SFB 631 and the
German Excellence Initiative via the Nanosystems Initiative Munich (NIM).

\end{document}